# Interband Plasmonics with p-block Elements

**Johann Toudert[*], Rosalia Serna**

*Instituto de Óptica, CSIC, Madrid, Spain*

*\*johann.toudert@gmail.com*

**Abstract** We investigate the origin of the near-ultraviolet - visible plasmonic properties of three elemental materials from the p-block: Bi, Sb and Ga, with the aim to achieve and exploit unconventional plasmonic effects beyond those provided by noble metals. At such aim, we review and analyze a broad range of optically-determined dielectric functions $\varepsilon = \varepsilon_1 + j\varepsilon_2$ of these elemental materials available in the literature, covering a wide photon energy range (from 0.03 to 24 eV). It is shown that the contribution of Drude-like carriers to $\varepsilon_1$ in the near-ultraviolet - visible is negligible for Bi and Sb and moderate for Ga. In contrast, the interband transitions of these elemental materials show a high oscillator strength that yields a strong negative contribution to $\varepsilon_1$ in the near-ultraviolet – visible. Therefore in these materials interband transitions are not a mere damping channel detrimental to plasmonic properties. Remarkably, these interband transitions *induce* fully (partially) the localized surface plasmon-like resonances taking place in Bi and Sb (Ga) nanostructures in the near-ultraviolet – visible. Plasmonic properties achieved through interband transitions, without Drude-like carrier excitation, are very attractive phenomena because they may give rise to a rich and broad class of nanostructures and metamaterials in which plasmonic effects will be tunable through the tailoring of band structure and by the occupancy of electronic states.

# 1. Introduction

Plasmonics has demonstrated to be one of the most exciting fields of nanotechnology, with possible applications in leading markets such as those of energy or biotechnology.[1-3] This interest is based on the ability of plasmonic nanostructures to enhance, filter, harvest, confine, guide or re-direction light with high efficiency at the nanoscale.[4]

The simplest plasmonic nanostructure that we can picture consists of a deeply subwavelength nanosphere with a photon energy-dependent dielectric function $\varepsilon(E) = \varepsilon_1(E) + j\varepsilon_2(E)$ embedded in a dielectric medium (permittivity $\varepsilon_m$). A resonance occurs for the nanosphere polarizability at the photon energy E for which the condition $\varepsilon_1(E) = -2\varepsilon_m$ is fulfiled, if $\varepsilon_2(E)$ is small enough. It is thus clear that this resonance, which is responsible of the *plasmonic properties of the nanosphere (such as near-field enhancement or enhanced light absorption)*, occurs in a spectral region that is primarily driven by the *dielectric function $\varepsilon$*. Note that the achievement of this resonance and the corresponding plasmonic properties *requires $\varepsilon_1(E)$ to be negative*, since $\varepsilon_m$ of a dielectric should be positive.

The physical origin of plasmonic properties has been conventionally attributed to the coupling of the material's free carriers to light. The response of these so-called "Drude carriers" is described by the Drude dielectric function, for which $\varepsilon_1$ decreases from 1 to markedly negative values when E decreases. Therefore for a nanosphere described by a Drude dielectric function, the nanosphere resonance condition is fulfiled *at a single photon energy*, at which a localized surface plasmon resonance occurs. This photon energy increases with the free carrier density/relative effective mass ratio.

The free carrier – based picture of plasmonic properties stands on the earliest known and most studied plasmonic materials: noble metals (Au and Ag). Au and Ag present a Drude-like dielectric function with a Drude carrier density/relative effective mass ratio N* in the $10^{22} - 10^{23}$ cm$^{-3}$ range, and *fulfil the nanosphere resonance condition at a single wavelength in the visible*.[5,6] Note that N* values of $10^{18}$ cm$^{-3}$ and above $10^{23}$ cm$^{-3}$ would make the resonance condition be fulfilled in the far-infrared and ultraviolet ranges, respectively.[7] Therefore in order to achieve plasmonic properties beyond the visible, a quest has started to identify *alternative plasmonic materials* presenting a dielectric function different from those of Au and Ag.[5-12] In addition novel enhanced plasmonic performances are sought including lower optical losses and novel physicochemical functionalities (e.g. switchable plasmonic properties).

Plasmonic properties were predicted early in non-noble elemental metals: alkaline, transition, and p-block metals.[6] The potential of these metals for applications is being currently revisited under the light of the newly acquired knowledge.[8-12] Upon adequate selection of the metal nature, localized plasmon resonances can be achieved in the near-ultraviolet, visible, or near-infrared. Al and In are particularly adequate for achieving sharp resonances and strong near-field enhancement in the deep and near-ultraviolet.[13-15] Furthermore the plasmonic properties of non-noble metallic compounds can be broadly tuned through their composition, as shown for titanium nitride-based compounds that are interesting alternatives to Au for applications in the visible.[9,16,17] Plasmonic properties can also be achieved for non-metallic compounds. Low optical losses and highly tunable plasmonic properties have been demonstrated in transparent conducting oxides.[9,18] Their plasmonic properties originate from the introduction of a controlled density of Drude carriers in the material's empty conduction band, though impurity doping or carrier transfer.[18] By such means, the range of plasmonic properties can be tuned from the far-infrared to the near-infrared, and switched dynamically. In a similar way, high impurity doping in semiconductors such as Si or Ge also permits to achieve plasmonic properties that can be driven from the far-infrared to the mid-infrared.[9]

All these examples show that plasmonic properties are achieved in a very broad range of materials, well beyond noble metals, and they suggest that plenty of plasmonic materials remain to be discovered. In this context, the link between plasmonic properties and Drude carriers has to be questioned. This is especially the case for some of the p-block elemental materials that present a complex electronic band structure and a dielectric function that cannot be represented by a Drude dielectric function, such as Bi, Sb and Ga. In particular in our group we have been recently attracted by the peculiar electronic and optical properties of Bi. Bi is a semi-metal for which plasmonic properties have been reported in both the mid-infrared[19] and the near-ultraviolet – visible (localized surface plasmon-like resonances in nanostructures).[20,21] This *"dual" plasmonic behaviour is not compatible with a Drude behaviour that imposes a single spectral region for the plasmonic properties*. A similar incompatibility arises for the much less studied semi-metal Sb, which presents plasmonic properties in the mid-infrared,[22] together with an evident metallic aspect that suggests plasmonic properties in the visible and a volume plasmon in the vacuum-ultraviolet that has been probed by electron energy loss spectroscopy.[23] Also particularly interesting is Ga whose nanostructures have been studied in detail recently.[12, 24-27] The most stable solid Ga phase (α-Ga) presents plasmonic properties in the near-ultraviolet – visible (localized surface plasmon-like resonances in nanostructures)[24] while being known as a molecular metal in which covalent and

metallic behavior coexist.[27-29]

Given the relevance that nanostructures made of these elemental materials have demonstrated at this point, it is necessary to achieve a deeper understanding of the origin of their plasmonic properties. First we will start by performing a comprehensive review of their dielectric functions available in the literature. We have gathered the data in in a broad photon energy range that covers the spectrum from the far-infrared to the vacuum-ultraviolet. Second we will present a comparative analysis of the data in relationship with the electronic band structure of the elemental materials. This analysis shows that the plasmonic properties observed in the near-ultraviolet - visible for Bi and Sb are fully induced by interband transitions. However Ga shows a more complex behavior, and we propose that its plasmonic properties in the near-ultraviolet – visible have an hybrid origin, i.e. they are induced by both interband transitions and Drude-like carriers.

## 2. Dielectric Functions of Bi, Sb and Ga

We will center our discussion on the reported dielectric functions that were obtained by optical spectroscopy. The data are very dispersed and varied. So far, the dielectric functions of Bi, Sb and Ga have been studied on samples synthesized by different methods (polished bulk crystals, thin films grown by evaporation, pulsed laser deposition…) and presenting different nanostructures (crystallinity, surface roughness, thickness). In particular the monocrystals present optical anisotropy, in relation with their anisotropic crystalline structure (Bi and Sb: rhomboedral; Ga: monoclinic) and energy band diagram.[30-32] This optical anisotropy is particularly marked for Ga. Moreover, in each report, the characterization has been realized in a limited photon energy range with a specific characterization method (reflectance or transmittance spectroscopy, spectroscopic ellipsometry, angles of incidence and polarization, measurement temperature and atmosphere differing from one work to another). In figure 1, we gather the results given in the collected reports. Despite a significant data scattering that appeals at a broadband characterization of model samples, especially for Bi and Ga, general trends can be drawn for the dielectric functions of Bi, Sb and Ga, showing that the three elemental materials share common optical features.

Upon decreasing the photon energy E from 24 eV (vacuum-ultraviolet) to 0.03 eV, (far-infrared): the $\varepsilon_1$ function first decreases to reach a local minimum in the visible – near-infrared, then increases until reaching a maximum, and finally abruptly decreases. Therefore we can define *two regions where $\varepsilon_1$ is negative and suitable to fulfil an optical resonance condition*: one located in the infrared (region I) and another in the near-ultraviolet - visible (region II). This behaviour *strongly departs from that of a simple Drude metal* for which $\varepsilon_1$ would *only decrease* upon decreasing the photon energy in the whole 24 eV – 0.03 eV range.

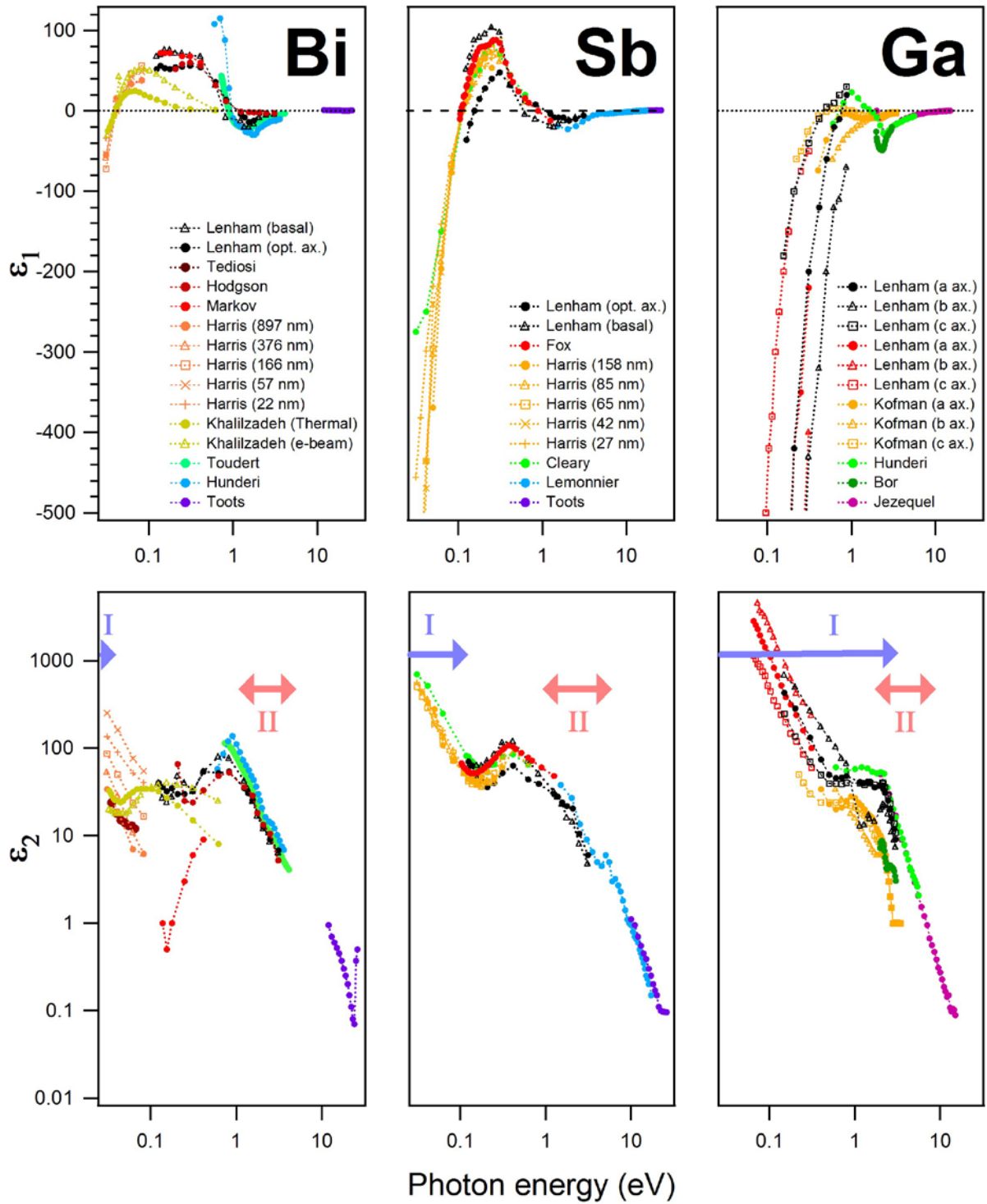

**Figure 1.** Optically-determined dielectric functions of solid Bi, Sb and Ga gathered from the litterature, real part $\varepsilon_1$ (top row) and imaginary part $\varepsilon_2$ (bottom row). The reported data have been obtained from the following references. Bi crystals: Lenham [ref. 37], Tediosi (290 K) [ref. 38], Hodgson [ref. 39], Markov [ref. 40]; Bi films: Harris [ref. 41], Khalilzadeh [ref. 19], Toudert [ref. 20], Hunderi (70 K) [ref. 30], Toots [ref. 42]. Sb crystals: Lenham [ref. 37]; Sb films: Fox [ref. 34], Harris [ref. 43], Cleary [ref. 22], Lemonnier [ref. 31], Toots [ref. 42]. Ga crystals: Lenham [ref. 35], Lenham [ref. 44], Kofman [ref. 45]; Ga films: Hunderi [ref. 46], Bor [ref. 47], Jezequel [ref. 48]. The regions where Drude-like carriers (region I) and interband transitions (region II) have a strong contribution to the dielectric function are depicted in the bottom row by blue and red arrows, respectively.

Instead, the Bi, Sb and Ga show rich dielectric functions that include both contributions from Drude-like carriers and interband transitions. The Drude-like carriers in these materials have been probed optically, yielding typical density/relative effective mass ratio N* (Bi: ref. 33, N* ~ $3 \times 10^{19}$ cm$^{-3}$, Sb: ref. 34, N* ~ $3 \times 10^{20}$ cm$^{-3}$, Ga: ref. 35, N* between $10^{22}$ and $2.10^{22}$ cm$^{-3}$, maximum in the basal a-b plane). From these values, it is expected that the Drude-like carriers in Bi, Sb and Ga induce a negative $\varepsilon_1$ up to a photon energy located in the far-infrared for Bi,[36] in the mid-infrared for Sb, and in the red part of the visible for Ga. The high photon energy side of region I in figure 1 follows nicely this trend, suggesting that the contribution of Drude-like carriers dominate the response of the three materials in this region.

In region II, the negative values of $\varepsilon_1$ are probably caused by interband transitions. These transitions are clearly evidenced in the $\varepsilon_2$ spectra of figure 1, that show a peak close to 0.8 eV for Bi, 0.3 eV for Sb, 2 eV for Ga, at slightly lower photon energies than the negative $\varepsilon_1$ minima. Such a behavior is typical of Kramers-Kronig consistency and the causality principle, which impose that transitions with high oscillator strength induce negative $\varepsilon_1$ values on their high energy side.[9,49]

## 2. Interband Origin of the Negative $\varepsilon_1$ in the Near-ultraviolet - Visible

In order to estimate the contribution of Drude-like carriers and interband transitions in region II, we have selected the most relevant dielectric functions for the three materials (for Bi ref. 30, for Sb refs 31 and 34 and for Ga ref. 46) and deconvoluted them using a linear sum of a Drude dielectric function and Kramers-Kronig consistent Lorentz oscillators accounting for interband transitions. We have chosen these dielectric functions as the most reliable because they show a good Kramers-Kronig consistency. Figure 2 shows the resulting deconvolution. It has been performed as follows: the parameters of the Drude dielectric function have been set to values for the Drude-like carrier density/relative effective mass ratio N* and for the collision frequency $\tau$ taken from the literature (see captions of figure 2), and the parameters of the Kramers-Kronig consistent oscillators were used as fit parameters.

It can be seen that the dielectric functions of Bi and Sb ("Exp" in figure 2) in region II are fully ruled by the Lorentz oscillators ("All Oscillators" in figure 2). The Drude function ("Drude" in figure 2) has a sizeable contribution only at much lower photon energies, in the infrared. Therefore, it is shown that the negative $\varepsilon_1$ and the plasmonic properties of Bi and Sb in the near-ultraviolet – visible region are *induced fully by interband transitions*.

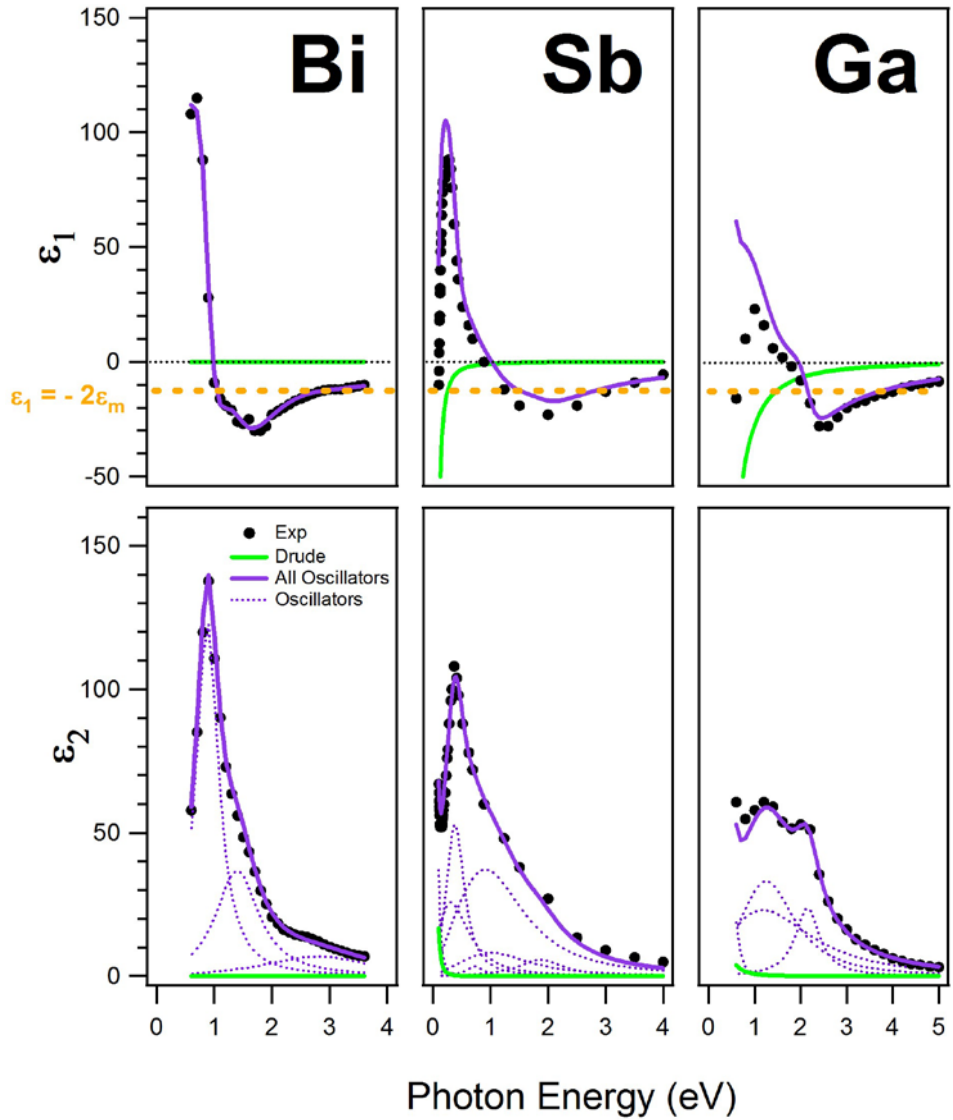

**Figure 2.** Deconvolution of the dielectric functions (black dots, "Exp") taken from ref. 30 (Bi), refs 31 and 34 (Sb) and ref. 46 (Ga). The deconvolution has been done using a sum of a Drude dielectric function (green lines) and Kramers-Kronig consistent Lorentz oscillators (purple dotted lines for individual oscillators ("Oscillators") and purple full lines for the sum of all the oscillators, "All Oscillators"). The parameters of the Lorentz oscillators where used as fit parameters whereas those of the Drude dielectric function were taken from the literature: Bi - $N^* \sim 3 \times 10^{19}$ cm$^{-3}$, $\tau = 300$ fs (ref. 33), Sb - $N^* \sim 3 \times 10^{20}$ cm$^{-3}$, $\tau = 31$ fs (ref. 34), Ga - $N^* = 2 \cdot 10^{22}$ cm$^{-3}$, $\tau = 21$ fs (ref. 35).

For Ga, both the Lorentz oscillators and Drude dielectric function play a significant role in region II. Therefore, both interband transitions and Drude-like carriers contribute to the negative $\varepsilon_1$ and plasmonic properties of Ga in the near-ultraviolet – visible, the main contribution being that of interband transitions. This observation is in line with the particular spectral shape of the $\varepsilon_2$ function on the high energy side of the 2 eV peak, which suggests that transitions between parallel bands dominate the response in the visible.[46] The important role of interband transitions in the visible is further supported by the electronic density of states of α-Ga, for which the Fermi level is located in a ~ 2 eV gap with a low sub-gap density of states.[27-29, 50] Note that the balance between the contributions of interband transitions and Drude-like carriers in the near-ultraviolet – visible may depend on the sample crystallinity (a lower collision frequency for the Drude-like carriers makes their contribution stronger in the visible). For an accurate quantification of these contributions, it will be necessary to perform a broadband characterization of a sample of interest.

## 3. Interband-induced Localized Surface Plasmon-like Resonances

Finally, we illustrate the contribution of interband transitions and Drude-like carriers on the near-ultraviolet – visible plasmonic properties of Bi, Sb and Ga by simulating the optical response of nanostructures made of these elemental materials. Figure 3 shows simulated spectra of the effective extinction coefficient of a medium consisting of nanospheres embedded in a transparent matrix. Simulations have been performed using several dielectric functions for the nanospheres: the best-fit dielectric functions to the literature data of figure 2 ("All Oscillators + Drude" in figure 3), the corresponding contribution of Drude-like carriers only ("Drude" in figure 3) and the corresponding contribution of interband transitions only ("All Oscillators" in figure 3).

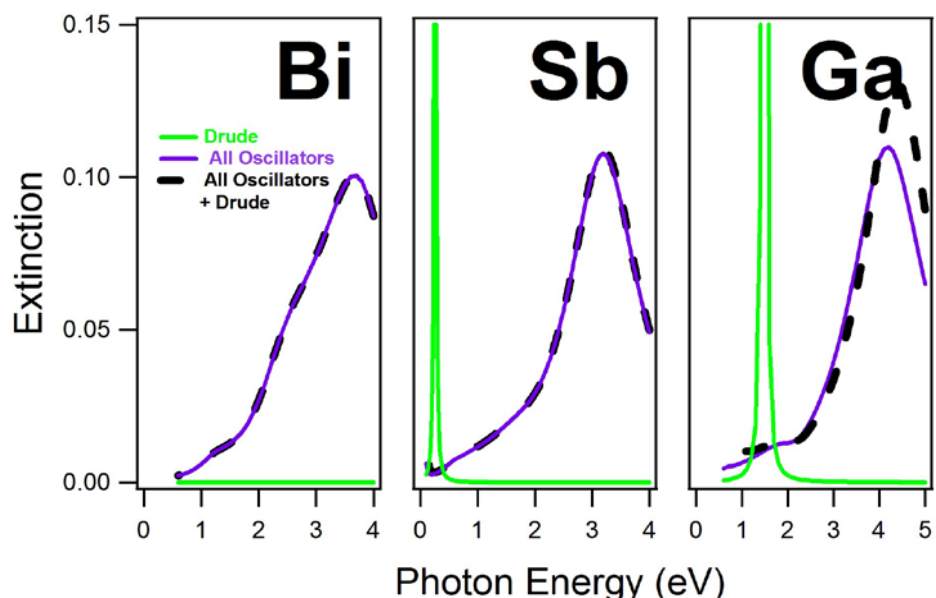

**Figure 3.** Simulated effective extinction coefficient spectra of a medium consisting of a low volume fraction (1%) of Bi, Sb, and Ga nanospheres embedded in a transparent matrix ($\varepsilon_m = 6.25$). The dielectric function of the nanosphere is: the best-fit dielectric function to the literature data (same as in figure 2, "All Oscillators + Drude", black dashed lines), the corresponding contribution of Drude-like carriers only (same as in figure 2, "Drude", green lines), the corresponding contribution of interband transitions only (same as in figure 2, "All Oscillators", purple lines).

Simulations done with the contribution of Drude-like carriers only show a sharp localized surface plasmon resonance, below 0.5 eV for Sb and around 1.5 eV for Ga confirming that the Drude-like carriers taken separately cannot induce a resonance in the near-ultraviolet – visible. In contrast, the simulations done with the contribution of interband transitions only are identical (for Bi and Sb) and very similar (for Ga) to those done with the best-fit dielectric functions to literature data that show a localized surface plasmon-like resonance with a maximum extinction in the near-ultraviolet. This supports the fact that the localized surface plasmon-like resonances of Bi and Sb in the near-ultraviolet - visible are induced by interband transitions only, whereas that of Ga is driven by interband transitions but with a significant but small contribution of Drude-like carriers. In that sense, we suggest that the resonances of Bi and Sb nanospheres should be named "interband-polaritonic resonances" and those of Ga nanospheres "hybrid interband-polaritonic/plasmonic resonances".

## 4. Interband Plasmonics: A New Paradigm in Photonics?

In the previous section we have demonstrated that the plasmonic properties of Bi and Sb in the near-ultraviolet - visible are *induced by* interband transitions, with no need of Drude-like carrier contribution. Interband transitions also play a key role in the case of Ga, where they *induce* the plasmonic-like properties together with the Drude-like carriers. This breaks the common belief that interband transitions are only a damping channel for plasmonic effects.[51,52] On the contrary, when strong enough, they may allow to generate or facilitate plasmonic properties. The fact that plasmonic effects can be generated by interband transitions, i.e. involving electron-hole pair excitation instead of free carriers, is interesting for many reasons. The demonstrated mechanism may help to understand and optimize the recently demonstrated photocatalytic properties of Bi nanoparticles[53,54] and more generally have implication for energy conversion schemes involving the collection of photogenerated carriers.[1] Furthermore, it opens the way to the design of novel nanostructures and metamaterials with a much higher tunability and switchability than those standing solely on noble metals, based on the tailoring of band structure and the dynamic control of band occupancy. Interband plasmonic properties seem to be achieved using other materials than Bi, Sb and Ga. In the p-block, Si and Ge show negative $\varepsilon_1$ values in the deep ultraviolet due excitonic transitions in these semiconductors. Similar effects are expected for compounds including p-block elements in the near-infrared.[55,56] Interband plasmonic properties in the visible have been evidenced early[57] and re-visited in organic materials,[58-60] and have been demonstrated very recently for topological insulators.[61] These are the first examples of a long list of materials suitable for achieving and exploiting unconventional plasmonic properties beyond those provided by noble metals. This shows the broad variety of possible building blocks for a new generation of switchable metamaterials for photonics.

## Acknowledgments

We acknowledge the Spanish Ministry for Economy and Competitiveness (Project AMELIE TEC 2012-38901-C02-01).